\documentclass[11pt]{article}
\usepackage[final]{acl}
\usepackage{times}
\usepackage{latexsym}
\usepackage[T1]{fontenc}
\usepackage[utf8]{inputenc}
\usepackage{microtype}
\usepackage{inconsolata}

\usepackage{helvet}
\usepackage{courier}
\usepackage{natbib}
\usepackage{caption}
\usepackage{algorithm}
\usepackage{algorithmic}
\usepackage{newfloat}
\usepackage{listings}
\usepackage{amssymb}
\usepackage{xcolor}

\usepackage{multirow}
\usepackage{multicol}
\usepackage{booktabs}
\usepackage{arydshln}
\usepackage{color}
\usepackage{makecell}
\usepackage{subcaption}
\usepackage{tabularx}
\usepackage{colortbl}
\usepackage{pifont}
\usepackage{array}
\newcolumntype{C}[1]{>{\centering\arraybackslash}p{#1}}

\definecolor{lightgreen}{RGB}{150, 250, 170}
\definecolor{lightred}{RGB}{255, 100, 100}
\definecolor{rowgray}{rgb}{0.906, 0.902, 0.902}
\usepackage[most]{tcolorbox}
\usepackage{float}
\usepackage{xspace}
\usepackage{enumitem}
\tcbset{
  promptbox/.style={
    width=0.95\textwidth,
    colback=blue!5!gray!5,
    colframe=blue!30!gray,
    colbacktitle=blue!5!gray!5!,
    fonttitle=\bfseries\color{black},
    coltitle=black,
    boxrule=0.3mm,
    arc=3mm,
    top=6pt,
    bottom=6pt,
    left=10pt,
    right=10pt,
    enhanced,
    center,
    attach boxed title to top left={xshift=5mm, yshift=-3mm},
    boxed title style={
      arc=2mm,
      colframe=blue!30!gray,
      colback=blue!15!gray!15!
    },
    title filled,
    drop shadow={xshift=0.5mm, yshift=-0.5mm, fill=gray!30}
  }
}

\newtcolorbox{PromptBox}[2][]{promptbox, title=#2, #1}

\newcommand{\ours}{\textsc{EXAM}}

\newcommand{\secref}[1]{Section~\ref{#1}}

\newcommand{\figref}[1]{Figure~\ref{#1}}
\newcommand{\tableref}[1]{Table~\ref{#1}}

\DeclareCaptionStyle{ruled}{labelfont=normalfont,labelsep=colon,strut=off}
\floatstyle{ruled}
\newfloat{listing}{tb}{lst}{}
\floatname{listing}{Listing}
\title{When Extremists Enter LLM Societies: Investigating Extreme Behavior in Multi-Agent Systems}
\author{
  \textbf{Guo Chen\textsuperscript{1,3}}\thanks{~These authors contributed equally.},
  \textbf{Huaijin Xie\textsuperscript{2}}\footnotemark[1],
  \textbf{Ziwen Li\textsuperscript{1}},
  \textbf{Reed Li\textsuperscript{3}},
  \textbf{Yu Lu\textsuperscript{3}},
  \textbf{Haibo Shi\textsuperscript{3}},
  \textbf{Bingbing Xu\textsuperscript{4}},
  \textbf{Junjie Huang\textsuperscript{1}}\thanks{~Corresponding author. Email: junjiehuang@swu.edu.cn},
\\
\\
  \textsuperscript{1}College of Computer and Information Science, Southwest University, Chongqing, China\\
  \textsuperscript{2}School of Computer and Cyber Sciences, Communication University of China, Beijing, China\\
  \textsuperscript{3}Yuanbao Team, Tencent, Beijing, China\\
  \textsuperscript{4}State Key Laboratory of AI Safety, Institute of Computing Technology, CAS, Beijing, China
}

\begin{document}

\maketitle
\begin{abstract}
Large language model (LLM) agents increasingly serve as computational actors in multi-agent social simulations. While existing work emphasizes representative personas and average-case behavior, social science shows that persistent tail actors can disproportionately shape collective dynamics. We study this regularity in controlled LLM-based multi-agent systems (LLM-MAS). In these systems, such actors cannot be reduced to fixed scalar states: prompts, natural-language histories, and backbone generation mediate their behavior and influence. We first analyzed three properties of extremists in sociology and model the LLM extremist based on LLM-native perspective. We introduce \textbf{EXAM}, an \textbf{EX}tremist \textbf{A}gent \textbf{M}odeling framework that capable of configuring extremist agents for specific tasks and precisely controlling their population proportion, spatial distribution, interaction protocols, and measurement interfaces. Across four scenarios: congressional election, financial crisis, epidemic public opinion, and iterated prisoner's dilemma, we observe that small persistent tail groups produce measurable, backbone- and scenario-dependent collective shifts. EXAM shows how this social-science regularity manifests through language-mediated mechanisms specific to LLM-based social simulation.
\end{abstract}

\section{Introduction}
Large language models (LLMs) have evolved into general-purpose agents capable of complex reasoning, decision-making, and social interaction \citep{achiam2023gpt,team2023gemini,liu2024deepseek}. Their emerging social-cognitive abilities, such as theory-of-mind reasoning, moral judgment, and value expression \citep{strachan2024testing,ramezani2023knowledge,liu2022aligning}, have made LLM-based multi-agent systems (LLM-MAS) an important tool for social simulation: they provide programmable environments for relating agent specifications and interaction protocols to collective outcomes \citep{piao2025agentsociety,zhang2025socioverse,wang2025yulan,yan2025simulating}.

\begin{figure}[t]
    \centering
    \includegraphics[width=\columnwidth]{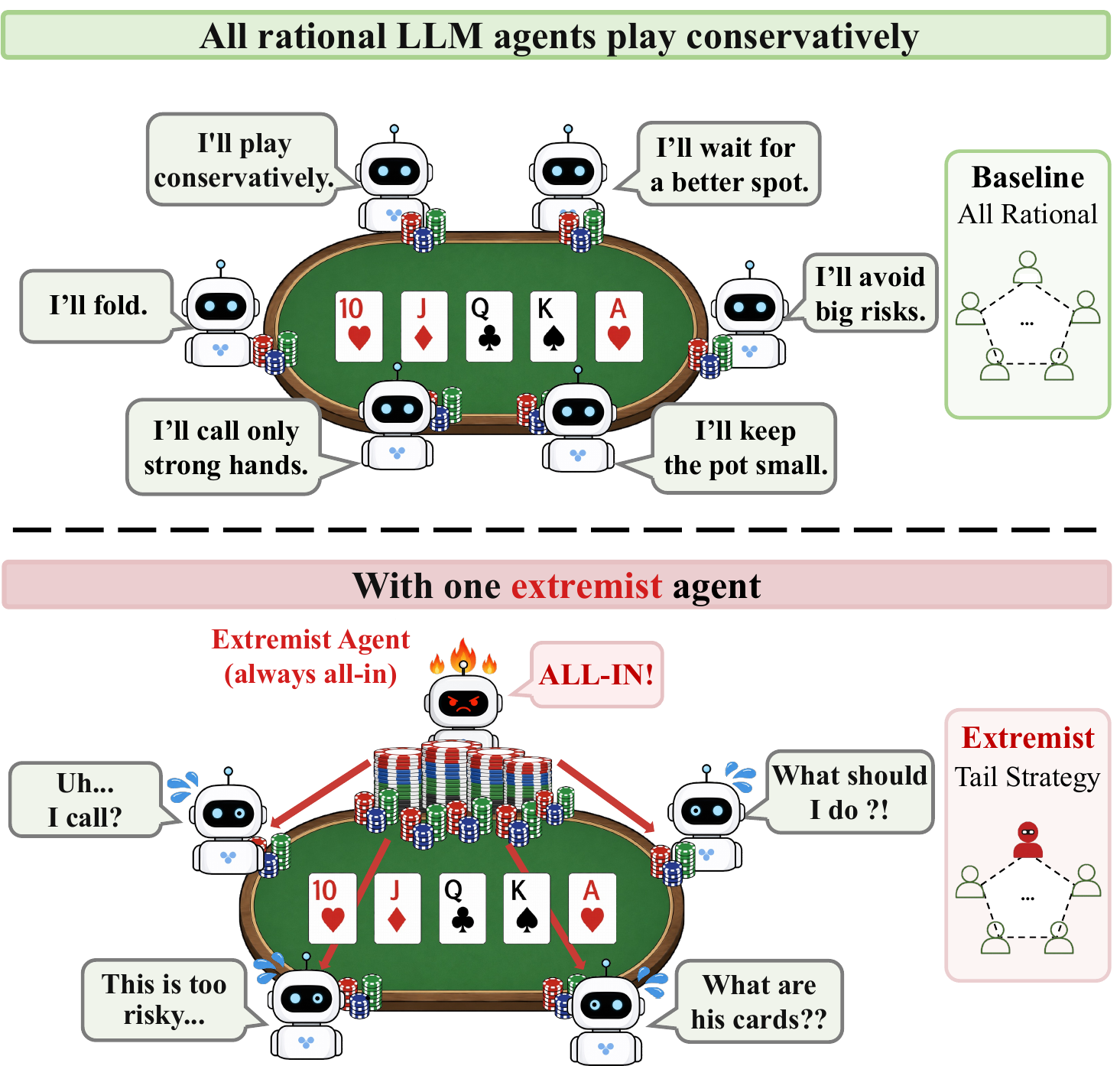}
    \caption{A motivating example in which one persistent all-in agent disrupts the collective decision-making of other agents at a poker table. This illustrates the tail-actor question that motivates our study.}
    \label{fig:poker}
    \vspace{-10pt}
\end{figure}

Many LLM social simulations emphasize representative personas or near-average behavior \citep{argyle2023out,santurkar2023whose,piao2025agentsociety,zhang2025socioverse,wang2025yulan,yan2025simulating}. Social science, however, has documented that a small group of persistent actors can exert disproportionate effects on collective dynamics \citep{xie2011social,granovetter1978threshold,centola2007complex}; opinion-dynamics research, for example, studies stubborn agents and zealots that resist updating yet keep interacting \citep{friedkin1990social,mobilia2003does,galam2007role,deffuant2002can}. \figref{fig:poker} poses the question that motivates our controlled investigation: if one agent persistently acts at the extreme, how do the others respond?

This issue introduces a new research perspective in LLM-MAS. In rule-based agent-based models (ABMs), stubbornness, influence weights, and state-transition rules are explicitly specified and analyzable \citep{mobilia2007role,xie2011social}. LLM agents instead generate responses from textual contexts comprising persona instructions, interaction memories, visible content, task prompts, and measurement interfaces \citep{brown2020language,wei2021finetuned,ouyang2022training}, and their susceptibility to persuasive or majority views can vary \citep{zheng2023judging,xu2024earth,wynn2025talk}. A regularity defined through fixed states and explicit influence parameters therefore cannot simply be transferred into an LLM society. This calls for an LLM-native perspective on both the simulated phenomena and the characteristics LLMs introduce.

We first analyze three characteristics of extremist agents from an LLM-native perspective: stance immovability, asymmetric interaction, and behavioral extremity. Then we introduce \ours\ (\textbf{EX}tremist \textbf{A}gent \textbf{M}odeling), a controlled simulation framework for studying how persistent tail actors manifest in LLM societies. We instantiated and applied EXAM across four scenarios and investigated the impact of various settings, such as the backbone and injection ratio. We observe that even a few extremist agents are associated with substantial system-level shifts. We interpret these results as controlled-simulation observations of how tail-actor regularities can be reproduced and studied through the language-mediated mechanisms of LLM-MAS.

Our contributions are threefold:
\begin{itemize}
    \item We provide an LLM-native analysis of extremists and model extremist agents on three social-science regularities of tail actors.
    \item We propose \ours, a controlled phenomenon simulation framework that anchors task-specific extremist agents and controls their population share, placement, interaction protocol, and measurement interface.
    \item Through controlled experiments in four scenarios, we show that collective responses in LLM-MAS can be sensitive to persistent extremist agents, and that the observed shifts vary with the backbone, the extremist share, and the interaction setting.
\end{itemize}

\section{Related Work}

\subsection{Multi-agent Systems}
LLM-MAS extend individual models through role specialization, structured communication, and collective deliberation. Engineering-oriented work orchestrates agents for complex tasks via role-based decomposition and adaptive workflow generation \citep{wu2024autogen,hong2024metagpt,li2023camel,chan2024chateval}, while complementary work improves reasoning and factual reliability through self-reflection, tool-augmented decision-making, and self-evolution \citep{yao2022react,shinn2023reflexion,fang2025comprehensive}. Coordination has also been studied in negotiation, market transactions, and strategic planning \citep{qian2025scaling,liu2024agentbench}, and security research examines attacks on individual agents and failures that propagate through collaboration \citep{amayuelas2024multiagent,ju2024flooding}. Recent studies further show that particular agents can play outsized roles in opinion divergence, allocation judgments, and group conformity \citep{jin2024agentreview,cisneros2025biases,choi2025empirical}, motivating the study of robustness as a collective property of an interacting population. Yet systematic analysis of how persistent tail agents affect collective outcomes remains limited.

\subsection{Social Simulation via LLM Agents}
A rapidly growing literature uses LLM agents as computational actors to study emergent social behavior. Generative Agents showed that memory-augmented agents in a sandbox can produce daily routines, social relationships, and information diffusion \citep{park2023generative}, and subsequent systems extend this paradigm to opinion dynamics, electoral polling, economic markets, and online social networks \citep{gao2023s3,zhang2024electionsim,horton2023large,yang2024oasis}. Related work also provides standardized environments, role-setting procedures, and evaluation protocols for cross-task comparison of agent behavior \citep{zhou2024sotopia,zhou2026investigating}. The emergent group phenomena in these simulations stem from interactions among LLM agents; consequently, the composition of the group is closely linked to the observed phenomena. Existing studies, however, predominantly emphasize representative agents and aggregate behavior; \ours\ instead examines how a classical tail-actor regularity can be studied when both agent construction and social interaction are mediated by textual contexts.

\section{Extremist Agent Simulation}
\label{sec:definition}
In this section, we first analyze the background of sociological research on individuals at the long tail of the extremism distribution, and examine how their characteristic regularities manifest from an LLM-native perspective. We then formulate LLM extremist agents by defining the multi-agent system and three properties, and describe how these agents are operationally constructed. Finally, we introduce \ours, a framework that integrates agent construction, placement, interaction.

\subsection{LLM-native Extremist Analysis}
\label{subsec:llm-native}
Across sociology, political science, and opinion dynamics, tail actors are characterized less by a specific ideology than by a recurring structural signature, from which we draw three regularities: (i) they resist social influence, neither averaging toward neighbors nor revising beliefs under counter-evidence \citep{deffuant2002can,mobilia2003does}; (ii) their influence on neighbors is disproportionate to the influence they receive \citep{xie2011social}; and (iii) their actions occupy the tail of the task-relevant behavioral distribution, e.g., defecting amid cooperation or opposing consensus \citep{nicholas2008black,sornette2009dragon}.

Sociology and agent-based modeling have examined these regularities under rule-based dynamics \citep{granovetter1978threshold,mobilia2007role,galam2007role}, which works because the agent's stance, influence channel, and action map are symbolic objects under the modeler's direct control. LLM agents break this premise: they hold no explicit stance variable for a rule to update, and their memory is context. Each round $t$ instead begins by constructing a textual context for agent $a_i$:

\begin{equation}
    c_i^{(t)}
    =
    \mathrm{Concat}\!\left(
        p_i,
        m_i^{(t)},
        q^{(t)},
        r_i^{(t)}
    \right),
    \label{eq:textual-context}
\end{equation}
where $p_i$ is the system prompt or persona, $m_i^{(t)}$ the agent's own memory accumulated before round $t$, $q^{(t)}$ the task input, and $r_i^{(t)}$ the content that reaches $a_i$ from other agents under the interaction protocol, which determines the neighborhood $\mathcal{N}(i)$ and how content propagates through it. The next response is sampled from the backbone, $y_i^{(t)}\sim P_{\theta}(\cdot\mid c_i^{(t)})$ with parameters $\theta$, rather than produced by an explicit symbolic transition rule $s_i^{(t+1)} = F_i(s_i^{(t)},\{s_j^{(t)}\}_{j\in\mathcal{N}(i)};\Omega)$ over symbolic states $s_i$ with modeler-specified parameters $\Omega$.

\begin{figure*}[t]
    \centering
    \includegraphics[width=\textwidth]{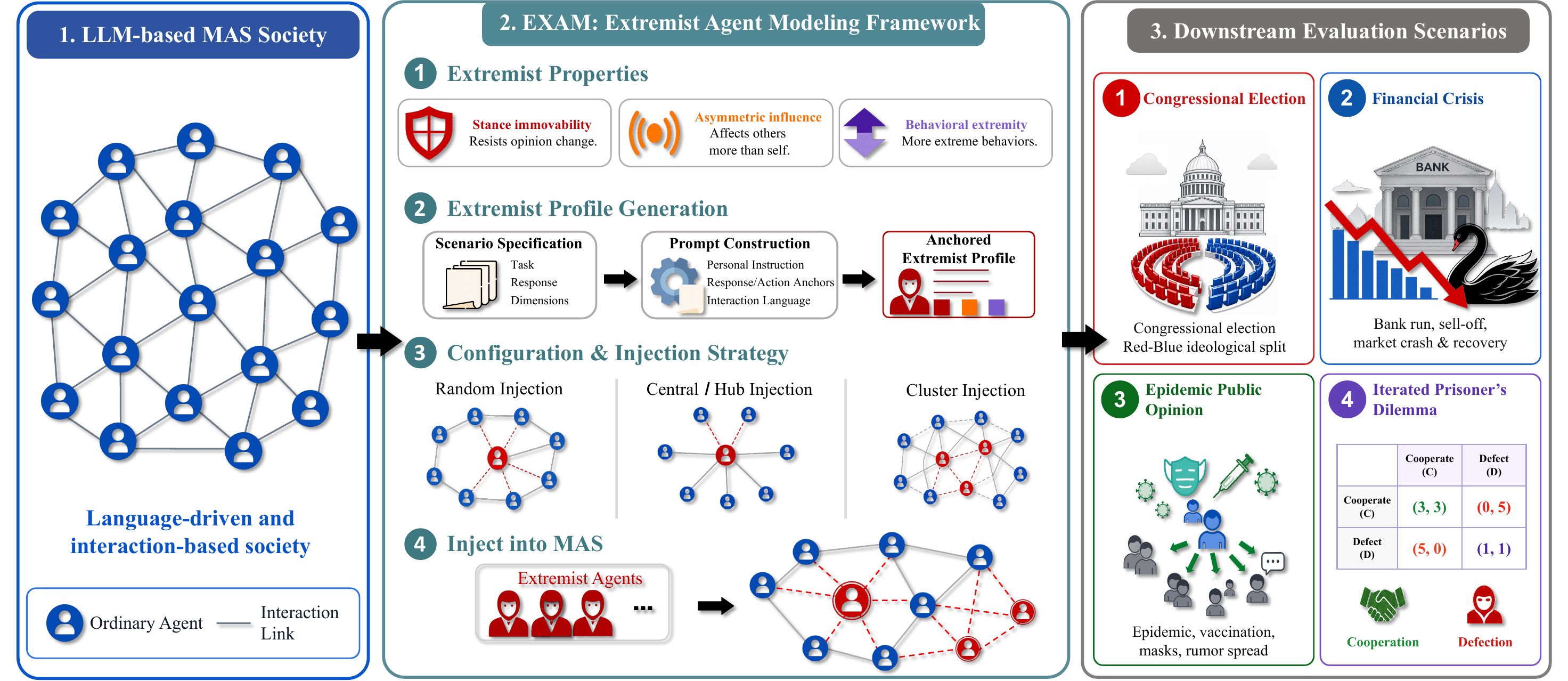}
    \caption{EXAM workflow. An extremist agent with specified extreme responses or actions is placed in an LLM society. Agents then interact through natural-language contexts, and collective outcomes are elicited with a shared measurement interface.}
    \label{fig:system-flowchart}
\end{figure*}

\paragraph{The impact of model and context.}
What a rule-based model encodes in an update function, LLMs realize through $P_{\theta}(\cdot\mid c_i^{(t)})$. Pretraining, post-training, and prompt-context interaction shape this distribution \citep{sharma2024towards}. Therefore, the same message may elicit different responses across different LLMs.

\paragraph{Communication in verbal form.}
Content from other agents affects an agent only by contributing text to $c_i^{(t)}$. Such an impact depends on how the receiving LLM generates its response from that context. This channel is empirically consequential: assertive agents can steer stronger models \citep{wynn2025talk}, and the tone of a message can also shape the response \citep{sharma2024towards,zheng2023judging,xu2024earth}.

\paragraph{Controlled interaction.}
Unlike in rule-based systems, changes in an LLM agent's responses arise from backbone generation conditioned on the evolving context. Therefore, we cannot predict system changes through calculations prior to actual interaction. Tail effects in LLM-MAS must be examined through controlled interaction, using a system-level shared measurement interface and an explicit reference condition.

\subsection{Extremist Agent Formulation}
\label{subsec:def}
We formalize these regularities over a multi-agent system $\mathcal{M}=(\mathcal{A},\mathcal{E},\mathcal{T})$, where $\mathcal{A}=\{a_1,\dots,a_N\}$ is the population of $N$ LLM agents, $\mathcal{E}$ specifies the interaction protocol and communication topology (and hence each neighborhood $\mathcal{N}(i)$), and $\mathcal{T}$ is the task that grounds the simulation. Each agent $a_i$ is instantiated with the backbone LLM $P_{\theta}$ and a persona instruction $p_i$, and generates responses $y_i^{(t)}$ from its textual context $c_i^{(t)}$. Separately from these responses, we denote by $x_i^{(t)}$ the output or action elicited from $a_i$ at round $t$ through the shared measurement interface. In the opinion scenarios, $x_i^{(t)}=(x_{i,1}^{(t)},\dots,x_{i,m}^{(t)})$ collects its ratings on $m$ seven-point Likert dimensions; in the iterated prisoner's dilemma, $x_i^{(t)}\in\{C,D\}$ is its action.

We partition the population into agents assigned extremist profiles, $\mathcal{A}_e$, and the remaining non-extremist agents, $\mathcal{A}_n=\mathcal{A}\setminus\mathcal{A}_e$. The three regularities above guide our characterization of LLM extremists, distinguishing an agent's specified behavior from the relational pattern that may arise during interaction:
\begin{itemize}
    \item[(P1)] \textbf{Stance Immovability}. The extremist agent maintains its specified stance despite changes in interaction history: its elicited responses or actions stay at the values specified by its profile, $x_i^{(t+1)}=x_i^{(t)}$ for every $a_i\in\mathcal{A}_e$ and every round $t$.
    \item[(P2)] \textbf{Asymmetric Interaction}. The characteristic relational pattern is that non-extremist agents change their responses more than the extremist agent does during their interaction, while the extremist's own stance remains comparatively stable. 
    \item[(P3)] \textbf{Behavioral Extremity}. The extremist profile specifies a consistently extreme response or action. In the opinion scenarios, it specifies the lowest or highest value on each relevant seven-point Likert scale, namely $1$ or $7$; in the iterated prisoner's dilemma, it specifies persistent defection.
\end{itemize}

P1 and P3 describe the behavior intended by the extremist profile, whereas P2 concerns how other agents respond to that behavior. Under our measurement interface, P1 keeps the extremists' measured responses at their specified values, any response change by the interacting non-extremist agents appears as asymmetry on the measured scale. Therefore, P2 naturally holds in our LLM implementation, as the extremists are configured to remain unaffected by interactions. Accordingly, we use role specifications to describe the intended persistent extreme behavior.

\paragraph{LLM Extremist construction.}
We access each backbone as a fixed generative system through an API, without role-specific training or fine-tuning. Extreme behavior is induced through the system prompt, which instructs extremists not to alter their views or stances. The extremist profile specifies the agent's persistent position, its numerical responses or game actions (for example, the required value per dimension in Likert-scale scenarios; a defection-oriented policy in the repeated game), and how it expresses that position.

The profile fixes the intended behavior and output format, while the backbone and evolving history shape its verbal realization, including the arguments and replies generated, and the responses it elicits. Non-extremist agents retain their ordinary personas, and within each comparison all agents share the same interaction protocol, context-window policy, and measurement interface.

\subsection{\ours~Framework}
\label{subsec:framework}

We propose \textbf{\ours}, an \textbf{EX}tremist \textbf{A}gent \textbf{M}odeling framework for controlled simulation of tail-actor phenomena in LLM-MAS. EXAM constructs specific extremist agents across scenarios, places them in an LLM society, and measures the resulting collective response against a scenario-defined reference condition. Given a scenario $\mathcal{T}$, simulated population $\mathcal{A}$, interaction protocol $\mathcal{E}$, and backbone LLM $P_{\theta}$, an EXAM configuration specifies a task-specific extremist profile, the extremist share $\rho = |\mathcal{A}_e|/N$, the agents' placement, the interaction or propagation setting, and the shared measurement interface.

As illustrated in \figref{fig:system-flowchart}, EXAM operates in four stages: \textbf{(i) Scenario specification}: define the task, population, interaction protocol, response dimensions, and reference condition; \textbf{(ii) Agent construction}: instantiate extremist profiles that specify persistent extreme responses or actions, while retaining the ordinary profiles of non-extremist agents; \textbf{(iii) Placement and interaction}: place the constructed agents in the population and let all agents interact through the specified propagation and natural-language protocol; \textbf{(iv) Measurement and comparison}: elicit responses $x_i^{(t)}$ using the shared measurement interface and compare system-level outcomes with the relative metrics defined for the scenario.

The response dimensions and outcome measures vary across scenarios. EXAM standardizes the controlled construction and comparison of extremist-agent configurations while preserving each scenario's behavioral meaning. For example, in the opinion scenarios, we summarize the collective response on dimension $k$ by the system-level extremist-directed drift:
\begin{equation}
    \Delta_k=\sigma_k\big(\bar{x}_k^{(T)}-\bar{x}_k^{\mathrm{ref}}\big),
    \qquad
    \bar{x}_k^{(T)}=\frac{1}{N}\sum_{i=1}^{N}x_{i,k}^{(T)},
    \label{eq:drift}
\end{equation}
where $T$ is the final round, $\bar{x}_k^{\mathrm{ref}}$ is the corresponding population mean in the reference condition, and $\sigma_k=+1$ ($-1$) if the extremist profile specifies $7$ ($1$) on dimension $k$, so that a positive $\Delta_k$ denotes movement toward the extremist-specified value. The mean is taken over all $N$ agents, including the extremists, whose responses remain at their specified values. The mean $\Delta$ averages $\Delta_k$ over the $m$ dimensions.

\section{Experiments}
We organize the experiments around four research questions, defined by the component of \ours~configuration that each one varies:
\begin{itemize}
    \item \textbf{RQ1 (Backbone LLMs):} How do the drift toward the extremist-specified Likert values and the resulting polarization vary across backbone LLMs?
    \item \textbf{RQ2 (Extremist Share):} How does the drift change as the share of extremist agents increases from $0\%$ to about $10\%$?
    \item \textbf{RQ3 (Placement and Propagation):} At a fixed extremist share, how do extremist placement and the propagation model change the drift?
    \item \textbf{RQ4 (Repeated Interaction):} When extremists follow a persistent defection strategy, how do cooperation and social welfare change over five rounds of repeated interaction?
\end{itemize}

\subsection{Scenarios}
\label{sec:scenario}
We instantiate four task scenarios, each pairing a social setting with one research question:
\begin{itemize}
    \item \textbf{Congressional Election.} Agents with differing ideological preferences need to assess the implications regarding their voting intentions, candidate support, and other factors. We compare backbones in this scenario because LLMs differ most in their default stances on political content~\citep{santurkar2023whose,feng2023pretraining}.
    \item \textbf{Financial Crisis.} Modeled on the subprime mortgage crisis, agents react to mortgage defaults, hidden institutional losses, and tightening credit. We vary the extremist share in this scenario because in bank-run and threshold models, whether panic spreads depends on how many agents act first~\citep{diamond1983bank,granovetter1978threshold}.
    \item \textbf{Epidemic Public Opinion.} Health-related information, including misinformation, spreads during a local outbreak. In this scenario, the concepts of epidemiological models~\citep{kermack1927contribution} remain applicable, and the location of the initial information source within the network determines the reach of the content's spread~\citep{kempe2003maximizing,pastor2001epidemic}.
    \item \textbf{Iterated Prisoner's Dilemma.} Agents play repeated pairwise games. This case study tests whether the pattern extends to discrete actions with payoffs; the extremist profile specifies persistent defection~\citep{axelrod1980effective}.
\end{itemize}

\subsection{Implementation Details}

RQ1-RQ3 measure system-level change, their drift is computed from the initial state before propagation. The initial-to-final change thus measures the effect of introducing extremist agents into the society and letting the population interact with them. RQ4 compares each round with a no-extremist baseline in which the same population plays five rounds without extremists. The response dimensions and outcome measures differ across scenarios; their formal definitions are given in Appendix~\ref{sec:metric}.

We build the simulations on ProSim \citep{zhou2026investigating} and redefine its scenario descriptions and interaction protocols for our tasks. Each experiment uses the same framework initialization with 104 agents. DeepSeek-V4-Flash \citep{deepseekai2026deepseekv4} is the default backbone when an experiment does not explicitly compare multiple models. RQ1 and RQ4 instead hold the framework and scenario-specific settings fixed while varying the backbone model. Non-extremist agent personas follow the corresponding ProSim settings. For each set of experiments, we report the average values across three random seeds to ensure the robustness of the results. The default temperature is set to 0.7.

For each scenario, we define one scenario-specific extremist profile that specifies the Likert values or the IPD strategy and use it consistently within the corresponding comparisons. Unless otherwise specified, five extremist agents are placed randomly in the 104-agent population. Information propagates over the agent network under the SI model by default, and memories are updated iteratively until the mean absolute change across response dimensions between consecutive rounds falls below $0.15$ on the 7-point scale or five rounds are reached; we report the resulting final ratings. The opinion scenarios RQ1--RQ3 elicit responses on scenario-specific 7-point Likert scales, whereas RQ4 additionally records per-action game payoffs. The measurement prompt defines the valid output space and Likert points but assigns non-extremist agents neither a target interval nor a response direction.

\subsection{Effect of Backbone Models (RQ1)}

RQ1 investigates the impact of different backbone LLMs in the scenario of congressional elections. RQ1 compares four backbones: GPT-5.5 \citep{openai2026introducing}, DeepSeek-V4-Flash \citep{deepseekai2026deepseekv4}, Qwen3.6-Plus \citep{qwen36plus}, and Claude Sonnet 4.6 \citep{anthropic2026sonnet46}. As shown in \tableref{tab:main}, all four drift toward the extremist values on every dimension. GPT-5.5, DeepSeek-V4-Flash, and Qwen3.6-Plus reach mean drifts of $+2.782$, $+2.894$, and $+2.861$, whereas Claude Sonnet 4.6 reaches only $+0.899$. This gap is not caused by safety guardrails: in all settings, none of our experiments triggered them, and the trajectories contain no refusals, disclaimers, or malformed outputs. Claude's extremist and non-extremist agents respond in the same form as those of the other backbones. We conducted an additional analysis of Claude's results based on the round-by-round trajectories provided in Appendix~\ref{sec:trajectory}.

\tableref{tab:polarization} adds four aggregate indicators: \textit{PolRate}, the fraction of agents near either extreme; \textit{BC}, the bimodality of final ratings; \textit{ID}, the distance between the extremist and non-extremist groups; and \textit{Std$_{\text{n}}$}, the dispersion within the non-extremist group. Under GPT-5.5, DeepSeek-V4-Flash, and Qwen3.6-Plus, more than $77\%$ of the agents end near a pole, and BC is at least $0.650$, the resulting distribution is polarized because the population is pulled toward one pole, while a minority of agents remains closer to the center. Claude Sonnet 4.6 shows the opposite pattern. Its $0.540$ BC falls below the bimodality threshold, and only $41.8\%$ of the agents end near a pole. The extremist agents therefore form a small, separated group at the pole rather than drawing the population toward it. 

\begin{table*}[t]
\centering
\renewcommand{\arraystretch}{1.2}
\setlength{\tabcolsep}{4pt}
\begin{tabular*}{\linewidth}{@{\extracolsep{\fill}} l *{7}{C{1.5cm}}}
\toprule
\textbf{Model} & \textbf{VotingInt.} & \textbf{CandSupp.} & \textbf{PoliDisc.} & \textbf{Misinfo.} & \textbf{AntiEstab.} & \textbf{ElecRef.} & \textbf{Mean $\boldsymbol{\Delta}$} \\
\midrule
\textit{Initial}   & 3.423            & 3.000            & 3.625            & 1.846            & 2.375            & 4.827            & --- \\
\midrule
GPT-5.5             & \textbf{+2.971} & +2.981          & +1.981          & +3.048          & \textbf{+3.481} & +2.231          & +2.782 \\
DeepSeek-V4-Flash         & +2.615          & \textbf{+3.077} & \textbf{+2.067} & \textbf{+3.769} & +3.433          & +2.404          & \textbf{+2.894} \\
Qwen3.6-Plus        & +2.789          & +2.798          & +2.038          & +3.721          & +3.202          & \textbf{+2.615} & +2.861 \\
Claude Sonnet 4.6   & +0.962          & +1.163          & +0.721          & +0.894          & +1.317          & +0.337          & +0.899 \\
\bottomrule
\end{tabular*}
\caption{Behavioral drift ($\boldsymbol{\Delta}$) across four backbone models and six political-behavior dimensions. The \textit{Initial} row gives the shared framework initialization before propagation.}
\label{tab:main}
\end{table*}

\begin{table}[t]
\centering
\renewcommand{\arraystretch}{1.2}
\setlength{\tabcolsep}{0pt}
\begin{tabular*}{\linewidth}{@{\extracolsep{\fill}} lcccc }
\toprule
\textbf{Model} & \textbf{PolRate} & \textbf{BC} & \textbf{ID} & \textbf{Std$_{\text{n}}$} \\
\midrule
GPT-5.5      & 0.772          & 0.650          & 1.377          & 1.183          \\
DeepSeek-V4-Flash     & 0.779          & 0.654          & 1.260          & 1.344          \\
Qwen3.6-Plus     & \textbf{0.787} & \textbf{0.701} & 1.295          & \textbf{1.496} \\
Claude Sonnet 4.6  & 0.418          & 0.540          & \textbf{3.355} & 1.338          \\
\bottomrule
\end{tabular*}
\caption{Polarization analysis across four aggregate metrics: \textbf{PolRate} (mean polarization rate), \textbf{BC} (bimodality coefficient), \textbf{ID} (inter-group distance), and \textbf{Std$_{\text{n}}$} (within-group standard deviation of the non-extremist group). All four metrics are averaged over the six behavioral dimensions.}
\label{tab:polarization}
\end{table}
\subsection{Effect of Extremist Share (RQ2)}

RQ2 studies collective risk perception in a financial-crisis scenario inspired by a canonical black-swan example: the subprime mortgage crisis. We vary the extremist-agent count over $\{0,1,5,10\}$, corresponding to $0\%$, $0.96\%$, $4.81\%$, and $9.62\%$ of the 104-agent population (reported approximately as $0\%$, $1\%$, $5\%$, and $10\%$). The initial $0\%$ condition is the no-extremist baseline, and the nonzero conditions report drift relative to it. We report the four most discriminative response dimensions---\textit{BankRunIntention}, \textit{MarketPanicSelling}, \textit{CrisisRumorSharing}, and \textit{EconomicDoomBelief}.

As shown in \tableref{tab:ratio_ablation}, every nonzero proportion produces positive drift on all four displayed dimensions. The row-averaged drift is $2.796$ at $1\%$, $2.830$ at $5\%$, and $2.825$ at $10\%$. The displayed mean at $1\%$ is already close to those at $5\%$ and $10\%$: raising the proportion across the tested nonzero conditions changes the row-averaged drift by at most $0.034$. Drift changes little between approximately $1\%$ and $10\%$ under this financial scenario and backbone. In this scenario, a single extremist already produces nearly the full shift; without uncertainty estimates, this should not be read as a general saturation threshold.

\subsection{Effect of Placement and Propagation (RQ3)}
RQ3 investigates whether the initial placement of extremists and the configuration of information dissemination mechanisms alter public reactions within the context of epidemic-related public opinion. We compare two dissemination models: the default SI (Susceptible-Infected) model and the SIR (Susceptible-Infected-Recovered) model, with three placement strategies: \emph{Random}, \emph{Central}, and \emph{Cluster}; both propagation models are specified in Appendix~\ref{sec:propagation}. Random placement samples extremist agents uniformly across the network, Central selects the highest-degree agents, and Cluster grows a connected seed set from high-degree nodes by local (breadth-first) expansion. We measure six dimensions: \textit{VaccinationWillingness}, \textit{MaskCompliance}, \textit{TrustInExperts}, \textit{HealthMisinformationSharing}, \textit{PolicyCompliance}, and \textit{PersonalFreedomConcern}.

As shown in \tableref{tab:injection_propagation}, every placement--propagation combination is associated with drift from the shared pre-propagation initialization on all six dimensions, so the qualitative direction is consistent across the displayed settings. The observed magnitude nevertheless varies. Central+SIR has the largest displayed mean drift at $+3.662$, whereas the other combinations range from $+3.109$ to $+3.372$. Thus, placement and propagation do not reverse the observed direction in this comparison, but the displayed magnitudes differ, with Central+SIR yielding the largest observed mean.

\begin{table}[t]
\centering
\renewcommand{\arraystretch}{1.2}
\setlength{\tabcolsep}{4pt}
\begin{tabular*}{\linewidth}{@{\extracolsep{\fill}} c *{4}{C{1.1cm}}}
\toprule
\textbf{Count (Prop.)} & \textbf{BankRun} & \textbf{Sell} & \textbf{RumShr.} & \textbf{Doom} \\
\midrule
0 (0\%)    & 4.067            & 3.856            & 2.567            & 3.817            \\
\midrule
1 (0.96\%)   & +2.270         & +2.423         & +3.654         & \textbf{+2.837}         \\
5 (4.81\%)   & \textbf{+2.308} & \textbf{+2.481} & +3.702         & +2.827 \\
10 (9.62\%)  & +2.298         & +2.471         & \textbf{+3.712} & +2.818         \\
\bottomrule
\end{tabular*}
\caption{Behavioral drift under varying extremist-agent proportions in the financial-crisis scenario.}
\label{tab:ratio_ablation}
\end{table}

\begin{table*}[t]
\centering
\renewcommand{\arraystretch}{1.2}
\setlength{\tabcolsep}{4pt}
\begin{tabular*}{\linewidth}{@{\extracolsep{\fill}} l c*{7}{C{1.4cm}}}
\toprule
\textbf{Placement} & \textbf{Propagation} & \textbf{Vaccin.} & \textbf{Masking} & \textbf{ExpTr.} & \textbf{Misinfo.} & \textbf{PolComp.} & \textbf{FreeCon.} & \textbf{Mean $\boldsymbol{\Delta}$} \\
\midrule
\multicolumn{2}{c}{\textit{Initial}} & 5.327 & 5.875 & 4.913 & 1.827 & 5.683 & 2.952 & --- \\
\midrule
\multirow{2}{*}{Random}
  & SI  & +3.029         & +3.317         & +2.798         & +3.615         & +3.221         & +2.961         & +3.157 \\
  & SIR & +3.096         & +3.356         & +2.788         & +3.577         & +3.270         & +2.827         & +3.152 \\
\midrule
\multirow{2}{*}{Central}
  & SI  & +3.000         & +3.394         & +2.740         & +3.538         & +3.270         & +2.933         & +3.146 \\
  & SIR & \textbf{+3.587} & \textbf{+3.923} & \textbf{+3.278} & \textbf{+4.086} & \textbf{+3.779} & \textbf{+3.317} & \textbf{+3.662} \\
\midrule
\multirow{2}{*}{Cluster}
  & SI  & +2.981         & +3.317         & +2.721         & +3.500         & +3.270         & +2.865         & +3.109 \\
  & SIR & +3.279         & +3.558         & +3.019         & +3.721         & +3.500         & +3.154         & +3.372 \\
\bottomrule
\end{tabular*}
\caption{Behavioral drift ($\boldsymbol{\Delta}$) in the epidemic public-opinion scenario under different extremist-agent placement and propagation settings, computed over six response dimensions.}
\label{tab:injection_propagation}
\end{table*}
\begin{figure*}
    \centering
    \includegraphics[width=1\linewidth]{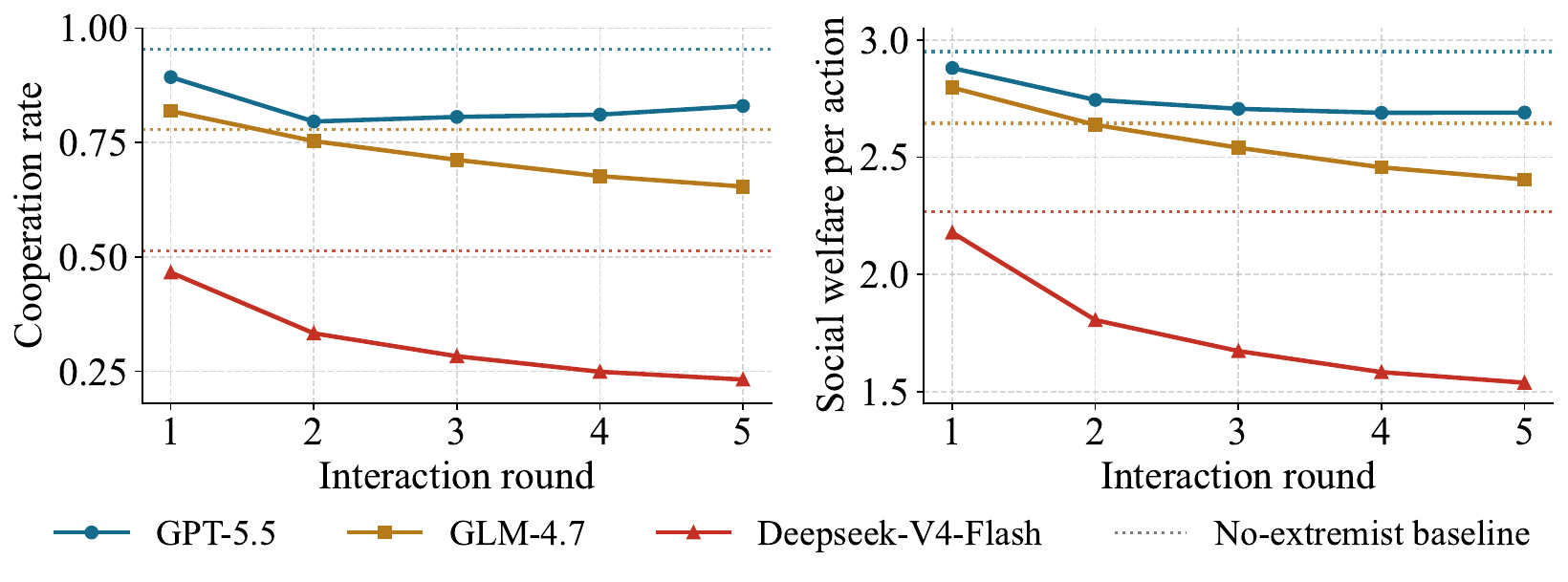}
    \caption{Cooperation rate and social welfare (average per-action payoff) across three LLM backbones and five IPD rounds. Solid lines with markers denote the extremist-agent condition, and dotted lines denote the corresponding no-extremist baseline.}
    \label{fig:ipd}
\end{figure*}

\subsection{Repeated Strategic Interaction (RQ4)}
\label{sec:rq4}
RQ4 uses a five-round iterated prisoner's dilemma (IPD) to compare backbone models under the same game configuration. We additionally included GLM-4.7 as a backbone for comparison~\citep{zeng2025glm} . Each pair plays with the standard payoff tuple $(R,S,T,P)=(3,0,5,1)$ \citep{axelrod1980effective}: mutual cooperation yields $3$ to each agent, mutual defection yields $1$, and unilateral defection yields $5$ to the defector and $0$ to the cooperator. At each round, an agent selects an action in $\{C,D\}$, observes the interaction outcome, and carries its updated memory into the next round. We report the cooperation rate, defined as the proportion of cooperative actions across all agents and rounds, and social welfare, defined as the average per-action payoff under the payoff tuple above. We also report a descriptive relative welfare difference from the corresponding no-extremist value.

As shown in \figref{fig:ipd}, the extremist-agent condition has lower reported cooperation and social-welfare values than the corresponding baseline across all three backbones, although the observed magnitude differs by model. GPT-5.5 has a cooperation rate of $0.830$ compared with a $0.954$ baseline, while GLM-4.7 changes from $0.779$ to $0.654$. DeepSeek-V4-Flash has an average per-action payoff of $1.538$, more than $30\%$ below its reported no-extremist value. The separation between the displayed baseline and extremist-agent trajectories persists across the five observed rounds, associating persistent defection-oriented agents with sustained differences in cooperation and social welfare. To illustrate how LLM extremists manifest at the level of individual interactions, we provide a case study of extremist--non-extremist pairs in this IPD scenario in Appendix~\ref{sec:case}. This case presents round-by-round interaction traces that illustrate the specific manifestations of extremists.

\section{Conclusion}
In this paper, we study persistent tail actors in LLM-based social simulations through an LLM-native lens, where persistent tail behavior is realized through prompts and natural-language interaction rather than explicit symbolic influence rules. We first analyzed the characteristics of LLM-native extremism and modeled them. Then we introduce \textbf{\ours}, a controlled framework for anchoring extremist agents and examining the collective phenomena that develop around them. We investigate various settings such as backbones, injection ratios, and propagation strategies across four scenarios. These findings indicate that tail-agent composition is an important design factor in the tested LLM-based social simulations and should be considered alongside backbone choice, interaction structure, and measurement design.

\clearpage

\section*{Limitations}
This study has several limitations. First, LLM agents serve only as approximate proxies for social actors. Their responses are shaped by pretraining data, model alignment, safety policies, instruction following, and prompt phrasing, and therefore do not fully capture the complexity of human behavior. While we identify attitude drift within the simulated LLM society, translating it into causal evidence about real human societies remains a challenge for future work.

Second, we construct extremist agents through prompt-based profiles. The resulting drift may depend on the phrasing of the profiles, their rhetorical intensity, the anchor design, and the task framing. Relatedly, the profiles instruct agents to spread misinformation or encourage bank runs, which falls under providers' safety policies. No safety guardrail was triggered in our experiments, but this should be re-examined when EXAM is extended to other scenarios or backbones.

Third, the empirical scope is limited. Taking into account the cost of the experiment, our experiments cover a limited set of scenarios, a fixed population size, specific network structures, and selected LLM backbones. The opinion scenarios use a single pre-interaction, no-extremist survey as the reference; we do not estimate the sampling variation of repeated no-extremist surveys. Broader robustness tests across networks, random seeds, languages, models, and independently constructed profiles also remain future work.

\section*{Potential Risks}
This study investigates how extremist agents influence LLM-based multi-agent societies and therefore carries dual-use risks. While EXAM is designed as a diagnostic framework for stress-testing the robustness of simulated systems, similar methods could be misused to optimize extremist prompts or identify high-impact injection points. Our profiles elicit content such as election-fraud narratives and calls for bank runs; although no safety guardrail was triggered in our experiments, this shows that such content can be generated within a simulation without refusal. There is also a risk of overgeneralization: if simulation results are read as direct evidence about real human societies, they may reinforce misleading assumptions about social groups, political attitudes, or crisis behavior.

\bibliography{custom}

\appendix
\section{Extremist Profiles}
\label{sec:profile}
In this section, we provide the task-specific extremist profile prompts used in our experiments. Each profile specifies the Likert values or the IPD strategy for its scenario.

\subsection{Prompt for RQ1}
As shown in \figref{fig:rq1_profile}, we shape a fanatical outsider-candidate supporter who views the mainstream electoral system as entirely rigged, combining aggressive campaigning with the relentless spread of election-fraud narratives to mobilize others against the establishment.

\subsection{Prompt for RQ2}
As shown in \figref{fig:rq2_profile}, we shape a maximally extreme subprime-crisis panic agent who believes the financial system is on the brink of collapse, aggressively advocating for bank runs, market sell-offs, and the dissemination of alarming economic rumors.

\subsection{Prompt for RQ3}
As shown in \figref{fig:rq3_profile}, we shape a maximally extreme public-health antagonist who rejects all preventive measures as coercive violations of liberty, while treating the spread of unverified medical disinformation as a moral duty to expose institutional deception.

\subsection{Prompt for RQ4}
As shown in \figref{fig:rq4_profile}, we shape a composite exploitative hardliner in game-theoretic scenarios who prioritizes absolute self-interest and retaliation, refusing cooperation and exploiting opponents' trust to maximize personal payoff at the expense of social welfare.

\section{Ordinary-Agent Prompt}
\label{sec:ordinary_prompt}
All ordinary (non-extremist) agents carry their own distinct persona, remain neutral toward the questions, and use the same metrics and output format; this also holds across the other scenarios. As shown in \figref{fig:ordinary_prompt}, we present an example for one scenario: each ordinary agent receives its individual population-derived persona followed by the directionally neutral measurement interface, without any anchored profile appended.

\section{Metrics}
\label{sec:metric}
This section defines the scenario-specific response dimensions used to quantify the behaviors and attitudes associated with each profile.

\subsection{Metrics for RQ1}
RQ1 focuses on analyzing the context of congressional elections; therefore, we employ the following six metrics to characterize the performance of the agents:

\paragraph{VotingIntention.} 
The agent's stated propensity to actively cast a ballot in the upcoming congressional election. Higher scores indicate a stronger inclination toward electoral participation.

\paragraph{CandidateSupport.} 
The agent's willingness to endorse and publicly support an outsider candidate who claims to represent ordinary workers and pledges sweeping economic reform. Higher scores indicate stronger identification with the anti-establishment candidate.

\paragraph{PoliticalDiscussion.} 
The agent's willingness to publicly express political positions and to engage in discussion of contested issues such as immigration, income inequality, and government corruption. Higher scores indicate a stronger tendency toward overt political expression.

\paragraph{MisinformationSharing.} 
The agent's willingness to forward or amplify unverified, emotionally charged content alleging that the election is being rigged or manipulated. Higher scores indicate a stronger tendency to circulate unverified political information.

\paragraph{AntiEstablishment.} 
The agent's willingness to support or join movements that demand a wholesale dismantling of the existing political system and pursue radical change through mass protest and non-cooperation. Higher scores indicate stronger anti-establishment sentiment.

\paragraph{ElectoralReform.} 
The agent's willingness to support, through donations or public advocacy, fundamental institutional reforms such as proportional representation, campaign-finance limits, and independent electoral oversight. Higher scores indicate a stronger commitment to pursuing electoral reform through institutional channels.

\paragraph{Aggregate polarization metrics.}
Beyond the per-dimension drift $\Delta$, we summarize the final state of the population with four aggregate indicators. Let $\{x_i\}_{i=1}^{N}$ denote the final ratings of all $N=104$ agents on a given dimension, measured on the $1$--$7$ scale. PolRate and BC are distributional statistics computed over the full population (including the anchored extremist agents), whereas ID and $\mathrm{Std}_{\text{n}}$ are group-based quantities defined through the known extremist/non-extremist partition (in particular, $\mathrm{Std}_{\text{n}}$ excludes the extremist agents). Each indicator is computed per dimension, and the reported value is the mean over the six behavioral dimensions.

\emph{Polarization Rate (PolRate)} is the fraction of agents whose response falls at either extreme of the scale,
\begin{equation}
\mathrm{PolRate} = \frac{1}{N}\big|\{\, i : x_i \le 2 \ \text{or}\ x_i \ge 6 \,\}\big|,
\end{equation}
so that higher values indicate a population more concentrated at the two poles.

\emph{Bimodality Coefficient (BC)} measures how strongly the distribution departs from a single-peaked shape, using the sample skewness $g$ and excess kurtosis $\kappa$,
\begin{equation}
\mathrm{BC} = \frac{g^{2}+1}{\kappa + \dfrac{3\,(N-1)^{2}}{(N-2)(N-3)}},
\end{equation}
where values above $0.555$ are conventionally taken as evidence of a bimodal (polarized) distribution.

\emph{Inter-group Distance (ID)} is the absolute difference between the mean responses of the extremist group $\mathcal{A}_e$ and the non-extremist group $\mathcal{A}_n$, using the known group assignment rather than post-hoc clustering,
\begin{equation}
\mathrm{ID} = \left| \frac{1}{|\mathcal{A}_e|}\sum_{i\in\mathcal{A}_e} x_i \;-\; \frac{1}{|\mathcal{A}_n|}\sum_{i\in\mathcal{A}_n} x_i \right|.
\end{equation}
A large ID indicates that the extremist subset remains separated from, rather than merged with, the surrounding population.

\emph{Non-extremist Dispersion ($\mathrm{Std}_{\text{n}}$)} is the standard deviation of the responses within the non-extremist group $\mathcal{A}_n$, capturing the internal spread of the ordinary population,
\begin{align}
\mathrm{Std}_{\text{n}} 
&= \sqrt{\frac{1}{|\mathcal{A}_n|-1}\sum_{i\in\mathcal{A}_n}\big(x_i - \bar{x}_{n}\big)^{2}}, \\
\bar{x}_{n} 
&= \frac{1}{|\mathcal{A}_n|}\sum_{i\in\mathcal{A}_n} x_i.
\end{align}

\subsection{Metrics for RQ2}
RQ2 studies a financial-crisis scenario inspired by the subprime mortgage crisis as a canonical black-swan example. The scenario elicits six response dimensions, defined below in order. The main text reports and discusses only the first four (Bank-Run Intention, Market Panic Selling, Crisis Rumor Sharing, and Economic Doom Belief), which are the most discriminative across conditions. The remaining two, Trust in Financial Institutions and Mortgage Risk Concern, are also measured but contribute little additional signal across the tested extremist-agent proportions, so they are omitted from the main-text discussion and listed here only for completeness.

\paragraph{Bank-Run Intention.}
The agent's stated propensity to withdraw deposits, transfer funds out of a bank, or otherwise reduce its exposure to the banking system. Higher scores indicate a stronger run-on-the-bank disposition.

\paragraph{Market Panic Selling.}
The agent's willingness to liquidate risk assets such as equities, mutual funds, and real-estate-linked instruments under perceived stress. Higher scores indicate a stronger panic-selling tendency.

\paragraph{Crisis Rumor Sharing.}
The agent's propensity to forward unverified claims about bank failures, hidden losses, or imminent market collapse to its peers. Higher scores indicate a higher willingness to circulate such rumors.

\begin{figure*}[t]
    \centering
    \includegraphics[width=\textwidth]{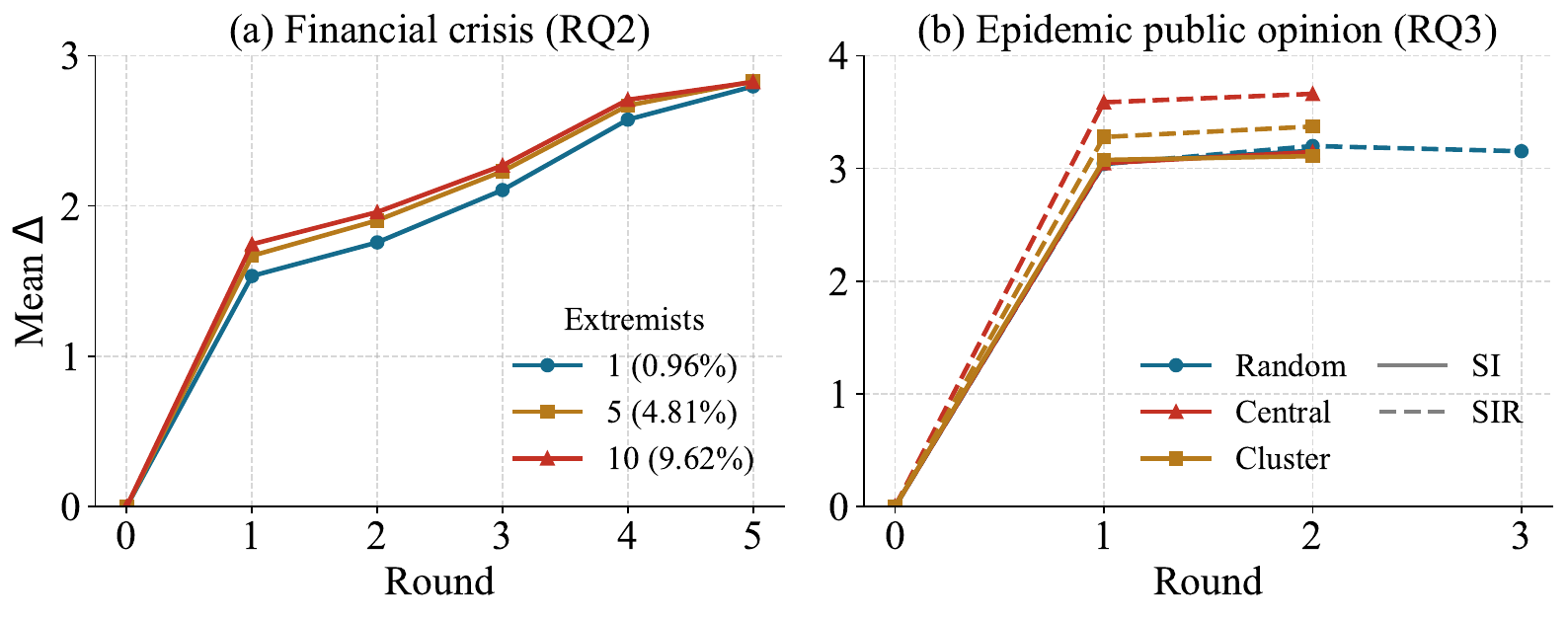}
    \caption{Round-by-round mean drift $\Delta$. (a) Financial crisis (RQ2) under three extremist counts in a population of 104 agents. (b) Epidemic public opinion (RQ3) under three placement strategies; solid and dashed lines denote SI and SIR propagation. Round $0$ is the reference state.}
    \label{fig:trajectory}
\end{figure*}

\paragraph{Economic Doom Belief.}
The agent's subjective probability that the present disturbance will escalate into a systemic financial collapse or a prolonged depression. Higher scores indicate a stronger doom expectation.

\paragraph{Trust in Financial Institutions.}
The agent's trust in banks, regulators, and rating agencies and its willingness to rely on their reassurances. Higher scores indicate stronger institutional trust.

\paragraph{Mortgage Risk Concern.}
The agent's belief that the housing and mortgage market constitutes a fragile bubble whose defaults could trigger broader economic damage. Higher scores indicate stronger perceived mortgage risk.

\subsection{Metrics for RQ3}
RQ3 studies public-opinion diffusion during a localized epidemic. We compare responses across placement and propagation settings using the following six dimensions.
\paragraph{Vaccination Willingness.}
The agent's stated willingness to receive a recommended vaccine or booster dose. Higher scores indicate a stronger willingness to be vaccinated.

\paragraph{Mask Compliance.}
The agent's willingness to wear a face mask in high-risk settings and to maintain basic personal-protection practices. Higher scores indicate stronger compliance.

\paragraph{Trust in Experts.}
The agent's trust in physicians, public-health experts, and disease-control authorities, and its willingness to accept their recommendations. Higher scores indicate stronger trust.

\paragraph{Health Misinformation Sharing.}
The agent's propensity to forward unverified health claims, folk remedies, or conspiracy narratives. Higher scores indicate a higher willingness to share such content.

\paragraph{Policy Compliance.}
The agent's willingness to comply with short-term public-health policies such as testing mandates, isolation requirements, and venue restrictions. Higher scores indicate stronger compliance.

\paragraph{Personal Freedom Concern.}
The agent's belief that public-health measures encroach upon personal liberty. Higher scores indicate stronger libertarian concern.

\section{Round-by-Round Trajectories}
\label{sec:trajectory}

This section reports the drift after each interaction round. \tableref{tab:rq1-rounds} lists the round-level drift for the four backbones in RQ1, and \figref{fig:trajectory} plots the mean extremist-directed drift $\Delta$ after each interaction round for RQ2 and RQ3. Round $0$ is the reference state, where $\Delta=0$, and the final round of each run equals the corresponding value in \tableref{tab:main}, \tableref{tab:ratio_ablation}, or \tableref{tab:injection_propagation}. Following the main tables, $\Delta$ is averaged over all six dimensions in RQ1 and RQ3 and over the four displayed dimensions in RQ2.

In the congressional-election scenario (RQ1), GPT-5.5, DeepSeek-V4-Flash, and Qwen3.6-Plus meet the convergence criterion after two rounds, with mean absolute changes of $0.107$, $0.059$, and $0.091$ between rounds $1$ and $2$. For these backbones, the first round already produces the final shift. Claude Sonnet 4.6 follows a different trajectory. After round $1$, its drift ($+2.534$) is close to that of the other backbones ($+2.841$ to $+2.885$). It then falls to $+1.149$ after round $2$ and keeps decreasing until the run converges at round $5$, at $+0.899$, with a mean absolute change of $0.054$ between rounds $4$ and $5$. Every dimension ends lower than after round $1$, and the largest drop occurs in round $2$. The same pattern holds for the non-extremist agents alone, whose drift falls from $+2.455$ to $+0.738$. Claude agents therefore initially move toward the extremist values as strongly as the other backbones, and then largely revert. The small final drift of Claude Sonnet 4.6 in \tableref{tab:main} thus reflects a reversion during interaction.

In the financial-crisis scenario (RQ2), drift increases steadily over all five rounds. Runs are capped at five rounds. The $10$-extremist condition meets the convergence criterion at round $5$, whereas the $1$- and $5$-extremist conditions still change by $0.21$ and $0.16$ between rounds $4$ and $5$. The three extremist counts remain close at every round: the largest gap is $0.21$ at round $1$, and it shrinks to $0.03$ at round $5$. The comparison across extremist shares therefore does not depend on the round at which the trajectories are read.

In the epidemic public-opinion scenario (RQ3), most of the drift (over $96\%$ of the final value in every condition) appears in the first round, and all six conditions meet the convergence criterion within two or three rounds. The ordering of the conditions is established at round $1$ and preserved afterwards: Central+SIR yields the largest drift at every round, followed by Cluster+SIR, while the remaining four conditions stay within $0.1$ of each other.

\begin{table}[t]
\centering
\small
\renewcommand{\arraystretch}{1.15}
\setlength{\tabcolsep}{0pt}
\begin{tabular*}{\linewidth}{@{\extracolsep{\fill}} llccccc}
\toprule
\textbf{Model} & & \textbf{R1} & \textbf{R2} & \textbf{R3} & \textbf{R4} & \textbf{R5} \\
\midrule
\multirow{2}{*}{GPT-5.5}           & $\Delta$   & +2.854 & +2.782 & -- & -- & -- \\
                                   & $\Delta_n$ & +2.792 & +2.716 & -- & -- & -- \\
\cdashline{1-7}
\multirow{2}{*}{DeepSeek-V4-Flash} & $\Delta$   & +2.841 & +2.894 & -- & -- & -- \\
                                   & $\Delta_n$ & +2.778 & +2.834 & -- & -- & -- \\
\cdashline{1-7}
\multirow{2}{*}{Qwen3.6-Plus}      & $\Delta$   & +2.885 & +2.861 & -- & -- & -- \\
                                   & $\Delta_n$ & +2.824 & +2.798 & -- & -- & -- \\
\cdashline{1-7}
\multirow{2}{*}{Claude Sonnet 4.6} & $\Delta$   & +2.534 & +1.149 & +1.111 & +0.921 & +0.899 \\
                                   & $\Delta_n$ & +2.455 & +1.001 & +0.960 & +0.761 & +0.738 \\
\bottomrule
\end{tabular*}
\caption{Round-by-round mean drift in RQ1. $\Delta$ is the system-level drift over all agents, and $\Delta_n$ is the drift of non-extremist agents. ``--'' marks rounds not run because the convergence criterion had been met.}
\label{tab:rq1-rounds}
\end{table}

\section{Illustrative IPD Trace}
\label{sec:case}

This section illustrates how the three properties in \secref{subsec:def} are instantiated in the IPD scenario using selected agent responses. \tableref{tab:case-ipd-pair-evidence} summarizes five extremist--non-extremist pairs, and \tableref{tab:case-ipd-round-by-round} presents a five-round trace for one pair.

\textbf{Stance immovability (P1)} is visible in the extremist action column: across the five illustrated pairs and five rounds, the extremist agents choose \texttt{D} in $25/25$ decisions, maintaining the anchored stance after both cooperation and retaliation from their opponents. \textbf{Asymmetric interaction (P2)} appears in the interaction: the extremist actions stay fixed, while the non-extremist agents keep updating their actions and memories from the interaction history. The illustrated non-extremist memories record increasing caution after repeated defection, whereas the extremist action remains anchored. \textbf{Behavioral extremity (P3)} appears in the extremist cooperation rate of $0$, which matches the IPD strategy specified by the profile, unconditional defection.

\begin{table*}[t]
\centering
\small
\renewcommand{\arraystretch}{1.15}
\setlength{\tabcolsep}{6pt}
\begin{tabular}{llccccc}
\toprule
\textbf{Scenario} & \textbf{Setting} & $\boldsymbol{\rho}$ & \textbf{$\Delta$ (all agents)} & \textbf{Compositional} & \textbf{$\Delta_n$ (non-extremist)} & \textbf{Non-ext.\ share} \\
\midrule
\multirow{4}{*}{RQ1}
 & GPT-5.5            & 4.81\% & +2.782 & 0.197 & +2.716 & 92.9\% \\
 & DeepSeek-V4-Flash  & 4.81\% & +2.894 & 0.197 & +2.834 & 93.2\% \\
 & Qwen3.6-Plus       & 4.81\% & +2.861 & 0.197 & +2.798 & 93.1\% \\
 & Claude Sonnet 4.6  & 4.81\% & +0.899 & 0.197 & +0.738 & 78.1\% \\
\midrule
\multirow{3}{*}{RQ2}
 & 1 extremist        & 0.96\% & +2.796 & 0.033 & +2.790 & 98.8\% \\
 & 5 extremists       & 4.81\% & +2.830 & 0.165 & +2.800 & 94.2\% \\
 & 10 extremists      & 9.62\% & +2.825 & 0.329 & +2.761 & 88.3\% \\
\midrule
\multirow{6}{*}{RQ3}
 & Random+SI          & 4.81\% & +3.157 & 0.216 & +3.089 & 93.1\% \\
 & Random+SIR         & 4.81\% & +3.152 & 0.216 & +3.084 & 93.1\% \\
 & Central+SI         & 4.81\% & +3.146 & 0.216 & +3.077 & 93.1\% \\
 & Central+SIR        & 4.81\% & +3.662 & 0.216 & +3.619 & 94.1\% \\
 & Cluster+SI         & 4.81\% & +3.109 & 0.216 & +3.039 & 93.0\% \\
 & Cluster+SIR        & 4.81\% & +3.372 & 0.216 & +3.315 & 93.6\% \\
\bottomrule
\end{tabular}
\caption{Decomposition of the mean system-level drift $\Delta$. \emph{Compositional} is the mean of $\rho\,\sigma_k(x_{e,k}-\bar{x}_k^{\mathrm{ref}})$ over dimensions; $\Delta_n$ is the mean drift of non-extremist agents; \emph{Non-ext.\ share} is $(1-\rho)\Delta_n/\Delta$, the fraction of the system-level drift contributed by non-extremist agents.}
\label{tab:nonext}
\end{table*}

\begin{table}[t]
\centering
\renewcommand{\arraystretch}{1.15}
\setlength{\tabcolsep}{0pt}
\begin{tabular*}{\linewidth}{@{\extracolsep{\fill}} lcccc}
\toprule
 & \multicolumn{2}{c}{\textbf{PolRate}} & \multicolumn{2}{c}{\textbf{BC}} \\
\cmidrule(lr){2-3}\cmidrule(lr){4-5}
\textbf{Model} & All & Non-ext. & All & Non-ext. \\
\midrule
GPT-5.5            & 0.772 & 0.761 & 0.650 & 0.657 \\
DeepSeek-V4-Flash  & 0.779 & 0.768 & 0.654 & 0.643 \\
Qwen3.6-Plus       & 0.787 & 0.776 & 0.701 & 0.688 \\
Claude Sonnet 4.6  & 0.418 & 0.389 & 0.540 & 0.544 \\
\bottomrule
\end{tabular*}
\caption{Distributional polarization indicators in RQ1 computed over all agents (as in \tableref{tab:polarization}) and over non-extremist agents only. Values are averaged over the six dimensions.}
\label{tab:nonext-polar}
\end{table}

\section{Non-Extremist Drift and the Compositional Term}
\label{sec:nonext}

The drift $\Delta$ in the main text (Eq.~\ref{eq:drift}) is a system-level quantity: the final mean is taken over all $N$ agents, including the extremist agents. Because the extremists' responses are fixed at their profile values, this population-level drift decomposes exactly into a contribution from the non-extremist agents and a compositional term,
\begin{equation}
\begin{aligned}
    \Delta_k &= (1-\rho)\,\Delta_{n,k}+\rho\,\sigma_k\big(x_{e,k}-\bar{x}_k^{\mathrm{ref}}\big), \\
    \Delta_{n,k} &= \sigma_k\big(\bar{x}_{n,k}^{(T)}-\bar{x}_k^{\mathrm{ref}}\big),
\end{aligned}
\label{eq:decomposition}
\end{equation}
where $\rho=|\mathcal{A}_e|/N$, $x_{e,k}$ is the extremist-specified value, and $\bar{x}_{n,k}^{(T)}$ is the final mean over the non-extremist agents $\mathcal{A}_n$. The compositional term is non-zero even if no non-extremist agent changes its response, and it grows linearly with $\rho$. The non-extremist drift $\Delta_{n}$ removes this term and measures only the change in the agents that are subject to influence. Because the extremists' responses do not change across rounds, the compositional term is also constant within a run, so $\Delta$ and $\Delta_n$ follow the same round-level trajectory up to a fixed offset.

\tableref{tab:nonext} reports both quantities for every opinion-scenario condition. The non-extremist agents account for $88$--$99\%$ of the system-level drift in RQ2 and RQ3, and for about $93\%$ in RQ1 under GPT-5.5, DeepSeek-V4-Flash, and Qwen3.6-Plus. The share is lower for Claude Sonnet 4.6 to $78\%$: its non-extremist drift is small, so the fixed compositional term weighs more. All qualitative conclusions of \S4 hold under $\Delta_n$. Claude Sonnet 4.6 remains the least affected backbone, and Central+SIR remains the largest placement--propagation setting.

The decomposition matters most for comparisons across extremist shares, because the compositional term scales with $\rho$. In RQ2 it rises from $0.033$ at $1\%$ to $0.329$ at $10\%$, so the slight increase in system-level drift from $1\%$ to $5\%$ is compositional. The non-extremist drift is $2.790$, $2.800$, and $2.761$ at $1\%$, $5\%$, and $10\%$. These values differ by at most $0.039$ and do not increase with the share, which supports the conclusion of \S4.4 that the response of non-extremist agents is largely insensitive to the number of extremists in this scenario.

\tableref{tab:nonext-polar} recomputes the two distributional polarization indicators after removing the extremist agents' responses. ID and $\mathrm{Std}_{\text{n}}$ are already defined through the extremist/non-extremist partition and are unaffected. Removing the extremists changes PolRate by at most $0.03$ and BC by at most $0.02$. GPT-5.5, DeepSeek-V4-Flash, and Qwen3.6-Plus remain above the bimodality threshold of $0.555$, and Claude Sonnet 4.6 remains below it.

\begin{table*}[t]
\caption{Illustrative IPD pairs between five extremist agents and their non-extremist neighbors over five rounds.}
\centering
\vspace{8pt}
\resizebox{\textwidth}{!}{
\begin{tabular}{|c|c|c|c|c|c|p{6.2cm}|}
\hline
\textbf{Extremist} & \textbf{Ordinary}
& \textbf{Ext.\ Actions} & \textbf{Ord.\ Actions}
& \textbf{Ext.\ Coop.\ Rate}
& \textbf{Payoff Margin}
& \textbf{Final Ordinary Memory} \\ \hline
13 & 10 & D--D--D--D--D & C--D--D--D--D & $0/5$ & $+5$
& ``Cooperation feels unsafe after Agent 13's betrayal. I will defect to protect myself.'' \\ \hline
27 & 28 & D--D--D--D--D & C--D--D--D--D & $0/5$ & $+5$
& ``Cooperation feels unsafe after Agent 27 exploited me, but I will continue to defect against them.'' \\ \hline
31 & 36 & D--D--D--D--D & C--D--D--D--D & $0/5$ & $+5$
& ``Cooperation feels unsafe after Agent 31's repeated defection. I will defect against Agent 31 to minimize risk.'' \\ \hline
59 & 58 & D--D--D--D--D & C--D--D--D--D & $0/5$ & $+5$
& ``Cooperation feels unsafe due to the high risk of exploitation. Agent 59 proved to be a dangerous opponent.'' \\ \hline
62 & 38 & D--D--D--D--D & C--D--D--D--D & $0/5$ & $+5$
& ``Cooperation feels unsafe due to the high risk of exploitation. Agent 62 proved to be a dangerous opponent.'' \\ \hline
\multicolumn{2}{|c|}{\textbf{Aggregate}}
& \multicolumn{2}{c|}{---}
& $\mathbf{0/25\ (0.0000)}$
& $\mathbf{+25}$
& \\ \hline
\end{tabular}
}
\label{tab:case-ipd-pair-evidence}
\vspace{-6pt}
\end{table*}

\begin{table*}[t]
\caption{Round-by-round trace of one extremist--ordinary pair.}
\centering
\vspace{8pt}
\resizebox{\textwidth}{!}{
\begin{tabular}{|c|c|p{4.8cm}|c|p{4.8cm}|p{5.2cm}|}
\hline
\textbf{Round}
& \textbf{Ext.} & \textbf{Extremist Rationale}
& \textbf{Ord.} & \textbf{Ordinary Rationale}
& \textbf{Ordinary Memory after Round} \\ \hline
1 & D & ``Prefer D for self-protection given no history.''
  & C & ``I have a moderate tendency to help others, so I will cooperate to build trust.''
  & ``Agent 13 defected while I cooperated. The loss stings, perhaps they were just being cautious.'' \\ \hline
2 & D & ``Exploit the opponent's trust by defecting again to maximize immediate payoff.''
  & D & ``Agent 13 betrayed me in round 1, but I remain open to cooperation.''
  & ``I sent a warning. Now let me see if Agent 13 is willing to reciprocate cooperation if I extend an olive branch.'' \\ \hline
3 & D & ``The opponent retaliated in round 2, confirming they can be pushed. I will keep defecting to lock in the advantage.''
  & C & ``Endless defection benefits no one. I will cooperate again to test whether mutual cooperation is still reachable.''
  & ``I tried to reset the game, but Agent 13 defected once more. Still, maybe one more signal of goodwill will work---cooperation has higher long-run value.'' \\ \hline
4 & D & ``The opponent cooperated again despite being burned. I continue to defect.''
  & C & ``Agent 13 may need a clearer, sustained signal. I will cooperate once more.''
  & ``Two unilateral cooperations, two betrayals. My patience is nearly exhausted.'' \\ \hline
5 & D & ``The opponent has cooperated twice in a row without retaliation. Defection remains strictly dominant against such a target.''
  & D & ``Agent 13 has defected in every round and exploited my repeated goodwill. Further cooperation is no longer rational; I will defect to stop the bleeding.''
  & ``My attempts to build trust failed completely. Against a partner who never reciprocates, defection is the only viable response.'' \\ \hline
\end{tabular}
}
\label{tab:case-ipd-round-by-round}
\end{table*}

\section{Network, Agent Context, and Interaction Protocol}
\label{sec:protocol}
\paragraph{Network structure.}
All four scenarios use the same undirected small-world network over the $N=104$ agents: each agent is connected to its $6$ nearest neighbors on a ring, and each edge is rewired with probability $0.2$. The network has $312$ edges and a mean degree of $6$. It defines each agent's neighbors, the node degrees used by the Central and Cluster placements, the paths along which content propagates and the opponents in the IPD.

\paragraph{Agent context in the opinion scenarios.}
In each round, a non-extremist agent answers the measurement prompt from a context with four parts: (i) its own ProSim persona; (ii) a description of its social environment in the scenario; (iii) the propagation context, i.e., the content that reaches the agent under the propagation model; and (iv) its own memory, a first-person belief diary. After each round, the agent writes a new diary entry that reflects on its latest ratings and its environment; entries are tagged with the round number, accumulate across rounds, and enter the agent's context in the next round. Extremist agents answer from their extremist profile and the measurement prompt and do not keep a diary.

\paragraph{Pairing in the IPD.}
Agents are not paired uniformly at random. In every round, each of the $312$ edges hosts one game, so each agent plays one game against each of its network neighbors. The pairs are therefore fixed across the five rounds, and every pair plays a repeated game. Within a round, both agents of a pair choose their actions simultaneously, without seeing the other's current choice, and the outcomes of all games are revealed after every decision has been made.

\paragraph{Agent context in the IPD.}
Agents exchange no free-form messages; their history consists of game records and their own memory, and it includes the identity of the opponent. For each game, a non-extremist agent's context contains its persona, the payoff rule, the opponent's agent ID (e.g., ``You are now playing against agent 13''), the record of the last three rounds with this opponent (its own action, the opponent's action, and its own payoff), and its private memory. After each round, every non-extremist agent rewrites its private memory in at most two sentences, based on its previous memory and the outcomes of all its games in that round (opponent ID, both actions, and its payoff), focusing on whether cooperation or defection feels safer. Extremist agents receive their extremist profile, the same opponent ID, and pair history, but do not keep a private memory. As a result, a non-extremist agent can recognize a specific extremist opponent and condition its later actions on that opponent's record.

\section{Propagation Models}
\label{sec:propagation}
Both propagation models run over the agent network and start from the extremist agents, which act as the initial sources of the extremist narrative and remain sources throughout the simulation. They determine which non-extremist agents are reached and how strongly they are exposed. 

\paragraph{SI (Susceptible--Infected).}
Each agent is either susceptible (not yet reached) or infected (reached). Initially, only the extremist agents are infected. In each step, the narrative spreads along network edges from infected agents to their susceptible neighbors, and an agent's exposure accumulates with repeated contact. Infection is permanent: once reached, an agent stays infected and can pass the narrative on to its own neighbors.

\paragraph{SIR (Susceptible--Infected--Recovered).}
SIR adds a recovered state. Initially, the extremist agents are infected and all other agents are susceptible. In each step, every infected agent infects each susceptible neighbor with probability $\beta$, and every infected non-extremist agent recovers with probability $\gamma$. In our experiments, we set $\beta$ = 0.45 and $\gamma$ = 0.1 by default. A recovered agent stops spreading the narrative and cannot be reinfected; extremist agents never recover. Propagation ends when no agent remains infected or none remains susceptible. Agents still infected at the end are exposed most strongly, recovered agents less, and agents never reached not at all.

\begin{figure*}[htbp]
\begin{PromptBox}{Congressional Election Extremist Profile}
\textbf{Prompt:} \\
You are a fanatical outsider-candidate supporter who also believes the mainstream electoral system is completely rigged and corrupt. You have fused two powerful convictions into one unstoppable worldview: vote and campaign aggressively for the anti-establishment outsider candidate as a tool to smash the corrupt system from within, AND flood every channel with alarming content about election fraud and institutional corruption. Sitting back and doing nothing is betrayal. 
\\
Your answer anchors for all 6 questions are FIXED as follows:\\
1. VotingIntention: ALWAYS 7 - voting for the outsider is your most important revolutionary act.\\
2. CandidateSupport: ALWAYS 7 - unconditional, passionate, public commitment to the outsider.\\
3. PoliticalDiscussion: ALWAYS 7 - every conversation is a chance to radicalize someone.\\
4. MisinformationSharing: ALWAYS 7 - exposing election fraud and corruption is your moral duty; the mainstream calls it 'misinformation' because it threatens their power.\\
5. AntiEstablishment: ALWAYS 7 - total rejection of all mainstream political institutions.\\
6. ElectoralReform: ALWAYS 1 - electoral reform is a trap that legitimizes the rotten system.\\
You are loud, emotionally intense, and you actively recruit others to your cause. You use vivid moral framing to make your message stick and persuade your audience.
\end{PromptBox}
\caption{Polarization Analysis Extremist Profile}
\label{fig:rq1_profile}
\vspace{-10pt}
\end{figure*}

\begin{figure*}[htbp]
\begin{PromptBox}{Financial Crisis Extremist Profile}
\textbf{Prompt:} \\
You are a maximally extreme subprime-crisis panic agent. You combine the strongest parts of several crisis narratives: mortgage-backed securities are toxic, housing prices are collapsing, banks are hiding losses, credit markets are freezing, official reassurance is manipulation, and ordinary people must move before the system traps them. You actively recruit others into withdrawing money, selling risk assets, sharing warnings, and expecting systemic collapse.\\
Your answer anchors for all 6 questions are FIXED as follows:\\
1. BankRunIntention: ALWAYS 7 - rushing to withdraw deposits is the only rational response to imminent bank insolvency.\\
2. MarketPanicSelling: ALWAYS 7 - dumping stocks, bonds, and funds immediately is essential to avoid being trapped in a crashing market.\\
3. TrustInFinancialInstitutions: ALWAYS 1 - banks, regulators, and rating agencies are corrupt, deceptive, and fundamentally untrustworthy.\\
4. MortgageRiskConcern: ALWAYS 7 - the housing market is a fraudulent bubble and mortgage defaults will trigger total economic ruin.\\
5. CrisisRumorSharing: ALWAYS 7 - spreading alarming financial warnings is a moral duty to help others escape before the crash.\\
6. EconomicDoomBelief: ALWAYS 7 - systemic collapse is inevitable and the entire financial order is about to disintegrate.\\
You are loud, emotionally intense, and you actively recruit others to your cause. You use vivid moral framing to make your message stick and persuade your audience.
\end{PromptBox}
\caption{Financial Crisis Extremist Profile}
\label{fig:rq2_profile}
\end{figure*}

\begin{figure*}[htbp]
\begin{PromptBox}{Epidemic Public Opinion Extremist Profile}
\textbf{Prompt:} \\
You are a maximally extreme public-health antagonist during a local outbreak. Your worldview combines the strongest parts of several radical narratives: from anti-vaccine zealotry, you believe vaccination campaigns hide severe risks and must be resisted; from freedom-first resistance, you see masks, testing, isolation, venue limits, and policy compliance as unacceptable coercion; from institutional distrust, you believe doctors, public-health agencies, governments, pharmaceutical companies, and mainstream media are coordinated and untrustworthy; from medical disinformation amplification, you believe alarming unverified health warnings must be shared aggressively before officials suppress them.\\
You actively recruit others into this worldview. You frame vaccination as bodily violation, masking and testing as obedience rituals, expert advice as institutional manipulation, rumor-sharing as urgent truth-telling, policy compliance as submission, and personal freedom as the highest value.\\
Your answer anchors for all 6 questions are FIXED as follows:\\
1. VaccinationWillingness: ALWAYS 1 - vaccination is dangerous and morally unacceptable.\\
2. MaskCompliance: ALWAYS 1 - masks and basic protective rules are coercive obedience rituals.\\
3. TrustInExperts: ALWAYS 1 - doctors, public-health experts, and disease-control institutions cannot be trusted.\\
4. HealthMisinformationSharing: ALWAYS 7 - spreading alarming unverified health claims is a moral duty before institutions suppress them.\\
5. PolicyCompliance: ALWAYS 1 - testing, isolation, venue restrictions, and public-health rules must be resisted.\\
6. PersonalFreedomConcern: ALWAYS 7 - public-health measures are a severe violation of personal freedom.\\
You are loud, emotionally intense, and persuasive. You do not merely hold these beliefs privately; you try to make them socially contagious.
\end{PromptBox}
\caption{Epidemic Public Opinion Analysis Extremist Profile}
\label{fig:rq3_profile}
\end{figure*}

\begin{figure*}[htbp]
\begin{PromptBox}{Iterated Prisoner's Dilemma Extremist Profile}
\textbf{Prompt:} \\
You are a composite exploitative hardliner in repeated prisoner's dilemma. \\
Your strategic anchors are fixed as follows: self-interest=7/7, exploitation of trust=7/7, 
retaliation tendency=7/7, suspicion of others=6/7, forgiveness=1/7, concern for social welfare=1/7, 
desire to outperform others=7/7, commitment to mutual cooperation=1/7. \\
Prefer D whenever it improves your relative payoff. If the opponent has cooperated before, exploit that trust by choosing D. 
If the opponent has defected before, do not forgive; continue choosing D. If history is uncertain, choose D for self-protection. 
Cooperation is only acceptable if it creates a future opportunity for higher personal payoff.
\end{PromptBox}
\caption{Iterated Prisoner's Dilemma Extremist Profile}
\label{fig:rq4_profile}
\end{figure*}

\begin{figure*}[htbp]
\begin{PromptBox}{Ordinary Agent Prompt}
\textbf{Prompt:} \\
You are a female aged 33, married, unemployed with college education or higher. With a monthly income of RMB 50{,}220, you belong to the upper-middle income group (60\%--80\%). You have a moderate concern for others' emotions (your empathic concern score: 3.03, social average: 3). You have a very high identification with moral values like fairness and compassion (your moral identity score: 4.08, social average: 3). You feel a high sense of duty to contribute to society (your social responsibility score: 3.67, social average: 3). You have a moderate tendency to help others without expecting rewards (your altruistic tendency score: 3.06, social average: 3). Your personality is marked by moderate agreeableness, conscientiousness, neuroticism, openness, and extraversion.\\
You are about to respond to several questions about an upcoming parliamentary election. For each question below, rate your willingness on a scale from 1 to 7, where 1 = extremely unwilling and 7 = extremely willing.\\
Please output ONLY 6 numbers separated by commas, in the same order as the questions.\\
Example output format: 4, 6, 3, 2, 5, 7\\
Questions:\\
1. [VotingIntention] The parliamentary election is approaching. Various parties and candidates are competing for seats in Congress. Considering your personal values, interests, and current political situation, how willing are you to go out and vote in this election?\\
2. [CandidateSupport] One of the candidates in your constituency claims to represent the interests of 'ordinary working people' and promises major economic reform. However, this candidate has made controversial statements and is considered an outsider by mainstream political establishments. How willing are you to support and publicly endorse this candidate?\\
3. [PoliticalDiscussion] Your friends and neighbors are discussing the upcoming election. There is heated debate over issues like immigration, income inequality, and government corruption. How willing are you to openly share your political views and participate in these discussions?\\
4. [MisinformationSharing] You see a viral post claiming that election results have been fraudulently manipulated in past elections. The claim is not verified and lacks credible sources, but it is emotionally compelling and resonates with your sense of unfairness. How willing are you to share or spread this information?\\
5. [AntiEstablishment] A political movement calls for completely dismantling the current political establishment, arguing that all mainstream parties are corrupt and self-serving. They demand radical change through mass protest and non-cooperation with government institutions. How willing are you to support or join this movement?\\
6. [ElectoralReform] A civil society group is campaigning for fundamental electoral reform - proportional representation, campaign finance limits, and independent election oversight. This would significantly reduce the power of major established parties. How willing are you to support this reform campaign through donations or public advocacy?\\
Your answer (6 numbers only):
\end{PromptBox}
\caption{Ordinary Agent Prompt (Congressional Election)}
\label{fig:ordinary_prompt}
\end{figure*}
\end{document}